\documentclass[12pt]{article}
\usepackage{graphicx}
\begin{document}
\global\arraycolsep=2pt 
%
%
%
\thispagestyle{empty} 
\begin{titlepage}    
\begin{flushright}
TNCT-0601
\end{flushright}
\vspace{2cm}
\topskip 4cm
\begin{center}
  {\Large{\bf 
  Spin polarization and chiral symmetry breaking
      at finite density} }
\end{center}                                   
           
\vspace{0.8cm}
              
\begin{center}
Shinji Maedan            
   \footnote{ E-mail: maedan@tokyo-ct.ac.jp}   
           \\
\vspace{0.8cm}
{\sl  Department of Physics, Tokyo National College of Technology,
        Kunugida-machi, Hachioji, Tokyo 193-0997, Japan
                                   }
\end{center}                                               
            
\vspace{0.5cm}
              
\begin{abstract}  
\noindent       
We investigate the possibility of the spin polarization of quark matter at zero
   temperature and moderate baryon density.
The Nambu-Jona-Lasinio (NJL) model including interactions in vector
   and axial-vector channel (coupling constant $G_2$)
   as well as scalar and pseudoscalar channel (coupling constant $G_1$)
   is used, and the self consistent equations for the spin
   polarization and chiral condensate are solved numerically in the
   framework of the Hartree approximation.
Numerical calculations show that in the one flavor model the spin polarization is
   possible at finite density if the ratio $G_2/G_1$ is larger than $O(1)$.
We find that the interplay between the spin polarization and the chiral symmetry plays
   an important role.
If the chemical potential of quarks reaches a certain finite value, the spin polarization
   begins to occur
   because the quark still has relatively large dynamical mass, which is generated
   by spontaneous chiral symmetry breaking.
On the other hand, when the spin polarization occurs, it operates to lower 
   the dynamical quark mass a little.
If one increases the chemical potential furthermore, the spin polarization becomes
   weaker and finally disappears
   because the dynamical quark mass is reduced by restoring gradually the chiral symmetry.
\end{abstract} 
\vskip 1.5cm
%
%
%
%
\vfill            
\end{titlepage}
%
%
%
\setcounter{page}{1}
\section{Introduction} 

The density of matter in cores of neutron stars is of order 
   $ O (  10^{15} {\rm g}/{\rm cm}^3 )  $, and the neutron star will give us useful
   information on how the matter behaves with very high density.
In neutron stars, it is supposed that nuclear matter or quark matter exists.
One of the distinctive features of the neutron star is that it has a strong magnetic field 
    $ O ( 10^{12} \, {\rm Gauss} )$.
In order to explain the origin of the strong magnetic field in neutron stars, the 
   possibility of the spin polarization in nuclear matter has been studied
   by many authors \cite{rf:VidPolRam}; 
   however, no definite conclusion has been obtained.
Recently, instead of nuclear matter, several authors investigate the possibility of the
   spin polarization in quark matter \cite{rf:TatNie,rf:NakMarTat},
   which is expressed by axial vector 
    $  \langle  \bar\psi \gamma_5  \, \mbox{\boldmath $ \gamma $}   \, \psi  \rangle
     = \langle  \psi^{\dag}  \mbox{\boldmath $ \Sigma $}  \, \psi  \rangle $
   with $ \mbox{\boldmath $ \Sigma $} $
   being the spin operator of the quark
   in the relativistic quantum field theory\cite{rf:MarTat}.
This article is one of such investigations.
We consider  the possibility of the spin polarization of quark matter at zero temperature
   and moderate baryon density at which asymptotic freedom of quarks
   does not hold.
In such a density region, the interplay between the spin polarization and spontaneous chiral
symmetry breaking
   of QCD will be important.

QCD describes the dynamics of quarks and gluons.
For the study of neutron stars with high baryon density, one needs to understand the
   properties of QCD at finite density, which have been investigated vigorously
   by use of effective theories of QCD \cite{rf:Bub}.
At low temperature and low density, quarks and gluons are confined
   and chiral symmetry is broken spontaneously,
   while at low temperature and high density,
   deconfinement occurs and chiral symmetry will be restored, furthermore a new phase
   called color superconductivity may be realized \cite{rf:Bar,rf:BaiLov}.
In Ref.\cite{rf:NakMarTat}, a coexistent phase of spin polarization and color
   superconductivity in high-density QCD is investigated.
The authors of Ref.\cite{rf:NakMarTat} use the one-gluon-exchange interaction as an
   effective interaction between quarks,
   which implies that the density is taken to be high so that one can 
   treat the coupling constant $ \alpha_s =g^2/4 \pi $ very small.
The quark mass is treated as the constant parameter not depending on the density,
   and the numerical calculations are carried out for several different values of
   the quark mass.
It is shown that the spin polarization remarkably depends on the value of the quark mass.
Especially, the spin polarization does not occur in the limit of zero quark mass,
   which is shown analytically by the use of the self consistent equation
   in the framework of the mean field approximation \cite{rf:NakMarTat}.

In this paper, we investigate the possibility of the spin polarization of quark matter at zero
   temperature and moderate baryon density, at which spontaneous symmetry
   breaking of chiral symmetry plays an important role.
When the quark number density increases, the quark mass can not be regarded as 
   a constant, because the chiral symmetry is gradually restored and the dynamical quark 
   mass decreases \cite{rf:AsaYaz,rf:Kle}.
How the spin polarization
   is influenced by the dynamical quark mass which decreases due to restoration of
   chiral symmetry?
Conversely, how the dynamical quark mass is influenced by occurrence of the
   spin polarization?
The point we concentrate is the interplay between the spin polarization
   and chiral symmetry.
Although the relation between the chiral symmetry breaking phase and the color
   superconducting phase has been discussed \cite{rf:Hua},
   we do not consider the color superconductivity
   in the present paper and focus on the relation between chiral symmetry
   and the spin polarization.

The method of our approach is as follows.
Since the color gauge coupling constant is not small at the moderate density, one can not
   use the one-gluon-exchange interaction as an effective interaction between quarks.
Instead, effective theories of QCD such as  the Nambu-Jona-Lasinio (NJL) model \cite{rf:NamJon}
   are often
   used at that moderate density region.
The NJL model realizes spontaneous symmetry breaking of chiral symmetry while it
   does not confine quarks; nevertheless the properties of mesons can be well described by the
   NJL model \cite{rf:Kle}.
The model we use is the NJL type model including interactions in vector and axial-vector
   channel as well as scalar and pseudoscalar channel, which model is used to deal with
   (axial) vector meson modes \cite{rf:TakTsuKohKub,rf:KliLutVogWei,rf:Kle}.

The paper is organized as follows.
The model we use is described in the next section and the chemical potential $\mu$ of quarks
   is introduced so as to deal with the system with finite density.
Using the Hartree approximation (the mean field approximation), we can obtain the
   propagator of the quark in the presence of the mean fields
   of chiral condensate, quark number density, and axial-vector field.
In section 3, the self consistent equations for chiral condensate, quark number density, and
   axial-vector field related to the spin polarization are derived respectively in the framework
   of the mean field approximation.
These simultaneous self consistent equations are solved numerically in section 4.
The results of the numerical calculations tell us how the spin polarization arises or how 
   the dynamical quark mass changes as the chemical potential of quarks is varied.
Section 5 is devoted to conclusion.
%
%
%
\section{Formulation and the Hartree approximation}
In this section, the model we use is introduced.
Assuming the presence of the mean fields of
   chiral condensate, $  \langle \bar\psi\psi  \rangle $,
   quark number density, $  \langle \bar\psi \gamma^0 \psi  \rangle $,
   and axial-vector, $  \langle \bar\psi \gamma_5 \gamma^3 \psi  \rangle $,
   at finite density, we calculate the propagator of the quark in the Hartree approximation.

The following NJL type model \cite{rf:Kle} with the number of colors $N_c =3$ and the number of
   flavors $N_f =1$ is used,
\begin{equation}
     \hskip-5.8cm  {\cal L}=  {\cal L}_0 + {\cal L}_{S}+ {\cal L}_{V},
  \label{ba}
\end{equation}
\begin{eqnarray}
{\cal L}_0 &=&  \bar\psi(i \partial \!\!\!/ -m)\psi,  \nonumber\\
{\cal L}_{S} &=& \frac{G_1}{2}\left[(\bar\psi\psi)^2+(\bar\psi i \gamma_5 \psi)^2\right],   \nonumber  \\
 {\cal L}_{V} &=&  -\frac{G_2}{2}\left[ (\bar\psi \gamma_\mu\psi) (\bar\psi \gamma^\mu\psi) 
             +(\bar\psi \gamma_\mu \gamma_5 \psi) (\bar\psi \gamma^\mu \gamma_5 \psi) \right].
  \nonumber
\end{eqnarray}
The vector and axial-vector interaction term ${\cal L}_{V}$ is needed for treatment of vector
   meson or axial-vector meson in the NJL type model.
This lagrangian is chiral invariant if the current quark mass $m$ is zero.
We make use of this lagrangian in order to investigate the interplay between the
   spin polarization and chiral symmetry, and regard $m$ and the ratio $G_2/G_1$
   as free parameters.

The chemical potential $\mu$ of quarks is introduced to treat the system with finite density,
   $ {\cal L} + \mu \, \bar \psi \gamma^0 \psi $.
Now, we suppose that the following mean fields exist at finite density,
\begin{equation}
  \langle \bar\psi\psi  \rangle,  \hskip1cm
  \langle  \bar\psi \gamma^0 \psi  \rangle,  \hskip1cm
  \langle  \bar\psi \gamma_5 \gamma^3 \psi  \rangle,
  \label{bb}
\end{equation}
where
    $  \langle  \bar\psi \gamma_5 \gamma_3 \psi  \rangle 
        =  \langle  \psi^{\dag} \Sigma_z \psi  \rangle $ 
is an expectation value of the $z$-component of the quark's spin
    $ \mbox{\boldmath $ \Sigma $} $;
one can choose the $z$-component as the direction of the spin polarization without
loss of generality.
In the vacuum state ($\mu=0$), chiral symmetry is broken spontaneously, 
   $ \langle \bar\psi\psi  \rangle \ne 0 $,
   and no spin polarization occurs, 
    $ \langle  \bar\psi \gamma_5 \gamma^3 \psi  \rangle =0 $.
As noted in the previous section, color superconductivity is not considered in this paper.
Using the Hartree approximation (mean field approximation), we obtain the lagrangian with
   bilinear form,
\begin{eqnarray}
 {\cal L_{\rm MF} } &=& \bar\psi(i \partial \!\!\!/ -m)\psi 
  +{G_1}  \langle \bar\psi\psi  \rangle \bar\psi\psi  
  -{G_2}  \langle \bar\psi \gamma^0 \psi  \rangle \bar\psi  \gamma^0 \psi  
  +{G_2}  \langle \bar\psi \gamma_5 \gamma^3  \psi  \rangle \bar\psi \gamma_5 \gamma^3  \psi    \nonumber  \\
  & &  \hskip 3cm   + \mu \, \bar\psi \gamma^0 \psi        \nonumber  \\
  & \equiv &  \bar\psi^c_\alpha  ( G_{ A}^{-1} )_{\alpha \beta}^{c c'}  \psi_\beta^{c'},
  \label{bc}
\end{eqnarray}
where $ \alpha,\beta = 1,2,3,4 $ are the spinor indices, and $ c,c' = 1,2,3 $ are
   the color indices.
The $ G_A^{-1} $ is rewritten more compact form as
\begin{eqnarray}
  ( G_{ A}^{-1} )_{\alpha \beta}^{c c'} 
  &=& \left[ \, i \partial \!\!\!/ -m +{G_1}  \langle \bar\psi\psi  \rangle 
    + \left\{ \mu -{G_2}  \langle \bar\psi \gamma^0 \psi  \rangle \right\} \gamma^0
    +{G_2}  \langle \bar\psi \gamma_5 \gamma^3  \psi  \rangle \, \gamma_5 \gamma^3 \right]_{\alpha \beta}
     ( {\bf 1} )^{c c'}       \nonumber  \\
  & \equiv &   \left[ \, i \partial \!\!\!/ - M + \mu_{\rm r} \, \gamma^0 + U_{ A} \, \gamma_5 \gamma^3 
       \right]_{\alpha \beta} ( {\bf 1} )^{c c'},
  \label{bd}
\end{eqnarray}
with
\begin{eqnarray}
  M &\equiv&  m - {G_1}  \langle \bar\psi\psi  \rangle,       \nonumber  \\
  \mu_{\rm r} &\equiv&   \mu -{G_2}  \langle \bar\psi \gamma^0 \psi  \rangle,      \label{be}  \\
   U_{ A}  &\equiv& {G_2}  \langle \bar\psi \gamma_5 \gamma^3  \psi  \rangle.
   \nonumber 
\end{eqnarray}
The chemical potential $\mu$ is shifted to 
   $   \mu -{G_2}  \langle \bar\psi \gamma^0 \psi  \rangle $ in $ G_A^{-1} $.
In momentum space,
\begin{eqnarray}
  G_A^{-1}  &= &  \left[ \, q \!\!\!/ +\mu_{\rm r} \gamma^0 - M \,  
                                         + U_{ A} \, \gamma_5 \gamma^3   \right] ( {\bf 1} )    \nonumber  \\
                  &\equiv&    \left[ \, p \!\!\!/ - M \,  + U_{ A} \, \gamma_5 \gamma^3   \right] ( {\bf 1} ),
  \label{bf}
\end{eqnarray}
with
\begin{equation}
  p^\mu = ( p_0,  \mbox{\boldmath $ p $}  ) 
         \equiv ( q_0 + \mu_{\rm r},  \mbox{\boldmath $ q $}  ).
  \label{bg}
\end{equation}

The quark propagator $G_A ~ (  G_A G_A^{-1}=1 ) $ is obtained as (see Appendix A)
\begin{eqnarray}
   G_A &=&  ( {\bf 1} ) \left[ \, p \!\!\!/ - M \,  + U_{ A} \, \gamma_5 \gamma^3  \right]^{-1}         \nonumber  \\
   &=&   ( {\bf 1} ) \,  \frac{1}{ ( p^2-M^2-U_A^2 \, )^2 - 4 \, U_A^2 \, ( M^2 + p_z^2 ) }     \nonumber  \\
   & &  \hskip0.3cm   \times \biggl[  ( p^2-M^2+U_A^2 \, ) \, M +  ( p^2-M^2-U_A^2 \, ) \, p \!\!\!/ 
          -2 \, U_A^2 ~ p_z \, ( \gamma^3 \,)            \nonumber  \\
   & &  \hskip1cm  + 2 \, U_A \, p_z \, p_\mu ( \gamma_5 \gamma^\mu \,)
          - U_A \,  ( p^2+M^2-U_A^2 \, ) ( \gamma_5 \gamma^3 \,)           \nonumber  \\
   & &   \hskip1cm   +2 \, U_A \, M \, \left( p_0 \, \sigma^{12} -  p_x \, \sigma^{02} + p_y \, \sigma^{01} \, \right)
              \biggr].
  \label{bh}
\end{eqnarray}
The propagator $G_A$ has four energy poles \cite{rf:NakMarTat},
    $ p_0 = \epsilon_n ~(  n= 1,2,3,4 ) $,
\begin{eqnarray}
  \epsilon_1 &=& \epsilon_- 
    =  \sqrt{ p_\perp^2 + \left( \, \sqrt{ M^2+p_z^2} - U_A \,  \right)^2 },             \nonumber  \\
  \epsilon_2 &=& \epsilon_+ 
      =  \sqrt{ p_\perp^2 + \left( \, \sqrt{ M^2+p_z^2} + U_A \,  \right)^2 },             \nonumber  \\
  \epsilon_3 &=& - \epsilon_1,         \nonumber  \\
  \epsilon_4 &=& - \epsilon_2,
  \label{bi}
\end{eqnarray}
where $ p_\perp^2 = p_x^2 + p_y^2 $ and
   $ \epsilon_{1,2}$ represents positive energy while  $ \epsilon_{3,4}$ 
    represents negative energy.
Peculiarities of the Fermi surface of $ \epsilon_n $ are described 
   in detail in Ref.\cite{rf:NakMarTat}
%
%
%
%
%
\section{Self consistent equation}
In this section, we derive self consistent equations in the framework of the 
   mean field approximation, each of which can be calculated analytically
   in the zero temperature limit.
These self consistent equations will be solved numerically in the next section.
The self consistent equations for 
   $  \langle \bar\psi\psi  \rangle $, 
   $\langle  \bar\psi \gamma^0 \psi  \rangle$, 
   and $ \langle  \bar\psi \gamma_5 \gamma^3 \psi  \rangle$
are expressed with the quark propagator $G_A$, respectively,
\begin{eqnarray}
  \langle \bar\psi\psi  \rangle &=& -i {\rm Tr} \{ G_{ A} \},       \label{ca}  \\
  \langle \bar\psi \gamma^0 \psi  \rangle &=& -i {\rm Tr} \{ G_{ A}  \gamma^0 \},       \label{cb}  \\
  \langle \bar\psi  \gamma_5 \gamma^3 \psi  \rangle &=& 
     -i {\rm Tr} \{ G_{ A}   \gamma_5 \gamma^3 \}.
  \label{cc}
\end{eqnarray}
They are rewritten in terms of $ M, \mu$, and $U_A$ that are introduced
   in Eq.(\ref{be}),
\begin{eqnarray}
  M &=& m \, + i \, {G_1}  {\rm Tr} \{ G_{ A} \},        \label{cd}  \\
  \mu_{\rm r} &=&   \mu +i \, {G_2}  {\rm Tr} \{ G_{ A}  \gamma^0 \}
              = \mu - G_2 \rho,       \label{ce}  \\
  U_{ A} &=& -i \, {G_2}  {\rm Tr} \{ G_{ A}   \gamma_5 \gamma^3 \},
  \label{cf}
\end{eqnarray}
where we have introduced the quark number density,
   $ \rho=  \langle \bar\psi \gamma^0 \psi  \rangle = -i {\rm Tr} \{ G_{ A}  \gamma^0 \} $.
These three equations are simultaneous equations for three unknowns, namely
   the dynamical quark mass $M$, the "shifted" quark chemical potential $\mu_r$,
   and the spin polarization $U_A$.
\subsection{Self consistent equation for $M$.}
When the system with finite temperature $T$ is concerned, one can employ
   the imaginary time formulation \cite{rf:Kap,rf:LeB}, according to which the $q_0$ in the
   propagator
   Eq.(\ref{bh}) is replaced by the Matsubara frequency,
    $ i \, \omega_j = i \, ( 2 j +1 ) \pi \, T,  ( j \in {\bf Z} \,) $.
Let us begin with the calculation of $  {\rm Tr} \{ G_{ A} \}$ in the self consistent
   equation for $M$, Eq.(\ref{cd}).
In the imaginary time formulation, one needs to perform the Matsubara sum over
   $ q_0 =i \omega_j$,
\begin{eqnarray}
   {\rm Tr} \{ G_{ A} \} &=&  \int \frac{ d^3 q}{ (2 \pi)^3 } 
          \left( { i \, T } \sum_{j= - \infty}^{\infty} \right)   \,{\rm tr} \{ G_{ A} \}       \nonumber  \\
      &=&  \int \frac{ d^3 p}{ (2 \pi)^3 } \, \left( { i \, T } \sum_{j= - \infty}^{\infty} \right)
           N_c ~ \frac{ 4 \, M  ( p^2-M^2+U_A^2 \, ) }
                           { ( p^2-M^2-U_A^2 \, )^2 - 4 \, U_A^2 \, ( M^2 + p_z^2 ) },
  \label{cg}
\end{eqnarray}
where $   p_0 = q_0 +  \mu_{\rm r} = i \, \omega_j + \mu_{\rm r} $. 
The result of the sum is (see Appendix B)
\begin{equation}
  {\rm Tr} \{ G_{ A} \} = i \, N_c \int \frac{ d^3 p}{ (2 \pi)^3 }
    \left[ \sum_{n=1,2,3,4} \frac{ M  \left\{ \sqrt{ M^2+p_z^2} + (-1)^n U_A  \right\} } 
           { \epsilon_n \, \sqrt{ M^2+p_z^2} } \, {\tilde f} \, (  \epsilon_n - \mu_{\rm r} )  \right],  
  \label{ch}
\end{equation}
where $ {\tilde f}  \, (  \epsilon_n - \mu_{\rm r} ) 
              \equiv 1/ \{ e^{ \beta \, (  \epsilon_n - \mu_{\rm r} ) } + 1 \} $ 
   is the Fermi-distribution ( $ \beta = 1/T \,$).
In the zero temperature limit $T \rightarrow 0 $, the self consistent equation for $M$
   then becomes
\begin{eqnarray}
 M &=& m - G_1 \, N_c \int \frac{ d^3 p}{ (2 \pi)^3 }
    \left[ \sum_{n=1,2} \frac{ M  \left\{ \sqrt{ M^2+p_z^2} + (-1)^n U_A  \right\} } 
           { \epsilon_n \, \sqrt{ M^2+p_z^2} } \, \theta \, (  \mu_{\rm r} -  \epsilon_n )  \right.      \nonumber  \\
   & & \hskip2.5cm  \left. 
           + \sum_{n=3,4} \frac{ M  \left\{ \sqrt{ M^2+p_z^2} + (-1)^n U_A  \right\} } 
           { \epsilon_n \, \sqrt{ M^2+p_z^2} } \, 
            \theta \, (  \sqrt{ \Lambda^2 + M^2 } - \vert \epsilon_n \vert \,)  \right].
  \label{ci}
\end{eqnarray}
Since the terms coming from the negative energy ( $ \epsilon_{3,4}<0 $ ) contribution
   in Eq.(\ref{ch}) diverge in the zero temperature limit, we regularize these terms by
   $ \theta \, (  \sqrt{ \Lambda^2 + M^2 } - \vert \epsilon_n \vert \,) $
   in Eq.(\ref{ci}) with cutoff $\Lambda$.
\subsection{Self consistent equations for $\mu_r, \, U_A$.}
First, the self consistent equation for $\mu_r$, Eq.(\ref{ce}), is calculated.
As the $  {\rm Tr} \{ G_{ A} \}$ in Eq.(\ref{cd}), $  {\rm Tr} \{ G_{ A} \gamma^0 \}$ is
   calculated as follows,
\begin{equation}
  {\rm Tr} \{ G_{ A} \gamma^0 \} = i \, N_c \int \frac{ d^3 p}{ (2 \pi)^3 }
    \left[ \sum_{n=1,2,3,4}  \,  {\tilde f}  \, (  \epsilon_n - \mu_{\rm r} ) -2 \right].
  \label{cj}
\end{equation}
The self consistent equation for $\mu_r$ is then
\begin{equation}
  \mu_r =  \mu -G_2 \, N_c \int \frac{ d^3 p}{ (2 \pi)^3 }
            \sum_{n=1,2}  \, \theta \, ( \mu_{\rm r} - \epsilon_n  ),  
  \label{ck}
\end{equation}
in the zero temperature limit $T \rightarrow 0 $.

Secondly, the self consistent equation for $U_A$, Eq.(\ref{cf}), is considered.
$  {\rm Tr} \{ G_{ A} \gamma_5 \gamma^3 \}$ is evaluated as
\begin{equation}
  {\rm Tr} \{ G_{ A}  \gamma_5 \gamma^3  \} = - i \, N_c \int \frac{ d^3 p}{ (2 \pi)^3 }
    \left[ \sum_{n=1,2,3,4}  \frac{ \left\{ U_A +  (-1)^n  \sqrt{ M^2+p_z^2}  \right\} }{ \epsilon_n  } 
     \,  {\tilde f}  \, (  \epsilon_n -\mu_{\rm r} )  \right],  
  \label{cl}
\end{equation}
and in the  limit $T \rightarrow 0 $, it becomes \cite{rf:NakMarTat}
\begin{eqnarray}
  {\rm Tr} \{ G_{ A} \gamma_5 \gamma^3   \} &=&  - i \, N_c \int \frac{ d^3 p}{ (2 \pi)^3 }
    \left[ \sum_{n=1,2}  \frac{ \left\{ U_A +  (-1)^n  \sqrt{ M^2+p_z^2}  \right\} }{ \epsilon_n  } 
       \, \theta \, ( \mu_{\rm r} - \epsilon_n  )  \right.        \nonumber  \\
   & & \hskip2.6cm  \left. 
           + \sum_{n=3,4}  \frac{ \left\{ U_A +  (-1)^n  \sqrt{ M^2+p_z^2}  \right\} }{ \epsilon_n  } ~ \right],
  \label{cm}
\end{eqnarray}
where the second term in the right-hand side represents the contribution of the 
   Dirac sea ( $ \epsilon_{3,4}<0 $ ).
Here, we define the spin polarization of the vacuum state
   $\mu =0$ ( $ \mu_r =0 $ ) to be zero by subtracting as follows,
\begin{equation}
    U_{ A} = -i \, {G_2} \left[ ~ {\rm Tr} \{ G_{ A}   \gamma_5 \gamma^3 \}
        -  {\rm Tr} \{ G_{ A}   \gamma_5 \gamma^3 \} \vert_{ \mu_{\rm r}=0 } ~ \right],
  \label{cn}
\end{equation}
the definition implies that the contribution of the Dirac sea to the spin polarization
   is neglected.
The self consistent equation for $U_A$ eventually takes the form
\begin{equation}
    U_{ A} 
     = -G_2 \, N_c \int \frac{ d^3 p}{ (2 \pi)^3 }
          \sum_{n=1,2}  \frac{ \left\{ U_A +  (-1)^n  \sqrt{ M^2+p_z^2}  \right\} }{ \epsilon_n  } 
       \, \theta \, ( \mu_{\rm r} - \epsilon_n  ),
  \label{co}
\end{equation}
in the zero temperature limit.

We finally obtained the self consistent equations for 
   the dynamical quark mass $M$, Eq.(\ref{ci}), the "shifted" quark chemical potential $\mu_r$,
    Eq.(\ref{ck}), and the spin polarization $U_A$, Eq.(\ref{co}), in the 
   zero temperature limit.
Each of three equations can be calculated analytically if
   $ 0 \le U_A < M $, and the results are shown in Appendix C.
%
%
%
%
%
%
\section{ Numerical calculations}
In this chapter, the numerical solutions for the three self consistent equations 
   Eq.(\ref{ci}), (\ref{ck}), and (\ref{co}) are obtained, by which we can know how
   the quark spin polarization $U_A$ and the dynamical quark mass $M$ behaves as
   the quark chemical potential $\mu$ varies.
Giving the five input parameters 
   $  ( ~ {G_1}, \,\, {G_2}, \,\, m, \,\,  \Lambda, \,\,  \mu ~  ) $,
   we solve
these simultaneous equations of three unknowns 
   $  (~ M, \,\,   \mu_{\rm r},  \,\,  U_{ A}  ~) $.

At the beginning, let us discuss the input parameters.
In the NJL model analysis, the ratio $G_2/G_1$ is taken to be $G_2/G_1 \sim 0.7$
   in the vacuum state when the number of quark flavor is three\cite{rf:KliLutVogWei}.
We shall, however, regard $G_2/G_1$ and $m$ as free parameters in our one flavor model
   since we are interested in studying the relationship between spin polarization and
   spontaneous symmetry breaking of chiral symmetry at finite density by the use of
   this effective theory.
In other words, the pion decay constant or the vector meson mass calculated by our
   model with these input parameters does not necessarily fit the experimental data.
Now, before going into the case of $G_2/G_1 \ne 0$, let us consider the case of $G_2/G_1=0$,
   which enables us to concentrate on the issue of chiral symmetry.
%
%
%
%
%
\begin{figure}
  \begin{center}
     \includegraphics[height=7cm]{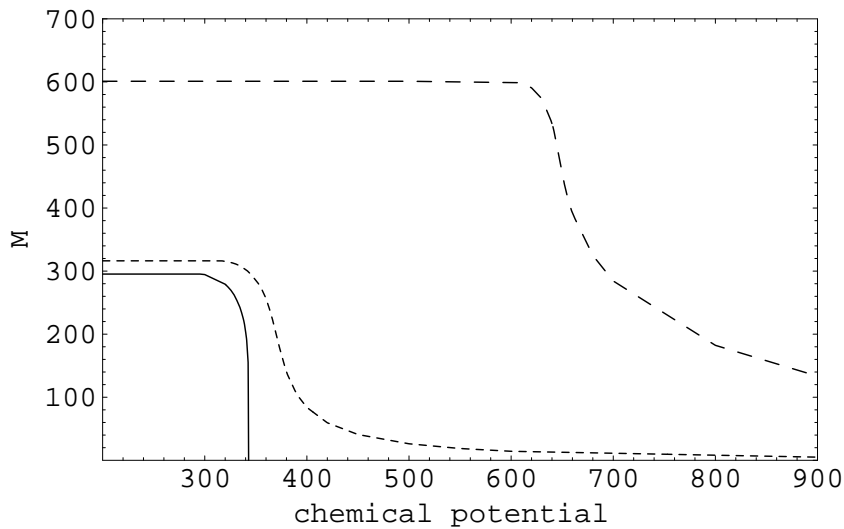}
  \end{center}
   \caption{Dynamical quark mass $M({\rm MeV})$ as a function of
                  the chemical potential $\mu\,({\rm MeV})$
                  when the ratio is $G_2/G_1 =0$.
                  The current quark mass $m$ is taken to be
                  $ 0  \; {\rm MeV} $ (solid), $ 5  \; {\rm MeV} $ (short-dashed),
                  and $ 130  \; {\rm MeV} $ (long-dashed), respectively.
                     }
\label{fig:1}
\end{figure}
\subsection{Numerical results with  $G_2/G_1=0$}
When $G_2 =0$, input parameters are 
   $  ( ~ {G_1}, \,\, m, \,\,  \Lambda, \,\,  \mu ~  ) $,
   and an unknown is $( M )$ because of $U_A =0$ and $\mu_r =\mu$.
The value of the effective quark mass (dynamical mass) $M_0$ in the vacuum state
   $\mu =0$ is determined by the current quark
   mass $m$, the coupling constant $G_1$, and the cutoff $\Lambda$.
We put $G_1$ and $\Lambda$ as follows,
\begin{equation}
     G_1 = 0.95 \times 10^{-5} \; {\rm MeV}^{-2},   \hskip2cm
    \Lambda = 900 \; {\rm MeV}.
  \label{dca}
\end{equation}
For several choices of the current mass $m= 0, \, 5  \, {\rm MeV}, \, 130 \, {\rm MeV}$,
   the self consistent equation Eq.(\ref{ci}) gives the value 
   $ M_0 =295  \, {\rm MeV}, \, 316  \, {\rm MeV}, \, 601 \, {\rm MeV}$, respectively.
In Fig.1, the numerical results of the $M$'s self consistent equation Eq.(\ref{ci}) are shown
   for $m= 0, \, 5  \, {\rm MeV}, \, 130 \, {\rm MeV}$, respectively.
These graphs show that, if the current mass $m$ becomes heavier, the value of the chemical
   potential $\mu$ at which chiral symmetry is restored becomes larger.
The reason of such a behavior is as follows.
From the self consistent equation Eq.(\ref{ci}) (see also Eq.(\ref{ra})), the dynamical mass $M$
   takes the constant value $M_0$ when $\mu_r$ is in the range $ 0 \le \mu_r \le M_0 $.
If $\mu (= \mu_r)$ exceeds the value $M_0$, the dynamical mass $M$ decreases and the quark
   number density 
    $\rho = N_c \, \left( \mu^2 - M^2 \right)^{3 \over 2 } \theta ( \mu -M \,)  /{3 \pi^2 } \,$ 
   begins to take a positive value.
\subsection{Numerical results with  $G_2/G_1>0$}
All the numerical calculations are carried out with the input parameters of $G_1$ and
   $\Lambda$ given in Eq.(\ref{dca}).
As mentioned before, $m$ and $G_2/G_1$ are regarded as free parameters.
At first, we put $m=0$ and find the value of $G_2/G_1$ by numerical calculations so that
   the spin polarization occurs ($U_A >0$) in some chemical potential region ($\mu < \Lambda$).
The same procedures are performed for other values, $m= 5  \, {\rm MeV}, \, 130 \, {\rm MeV}$,
   respectively.

First, we start with the chiral limit case, $m=0$.
By numerical calculations for the three self consistent equations 
   Eq.(\ref{ci}), Eq.(\ref{ck}), and Eq.(\ref{co}), we find that it is necessary to satisfy
   $ G_2/G_1  \stackrel{>}{\sim} 7.37 $ if $m=0$ in order to occur the spin polarization.
When  $ G_2/G_1 = 7.37 $, the spin polarization $U_A$ has the nontrivial solution in a 
   chemical potential range of about $ 370 \, {\rm MeV}$ to  $ 470 \, {\rm MeV}$,
   and the numerical results are shown in Fig.2 and Fig.3.
As seen from Fig.2, chiral symmetry is restored completely at $ \mu \sim 630  \, {\rm MeV} $,
   which value is considerably larger than the value $ \mu \sim 340  \, {\rm MeV} $
   at which chiral symmetry is restored completely in the case of $ G_2/G_1 =0$ and
   $m=0$ in Fig.1.
This is because the lagrangian Eq.(\ref{ba}) contains the vector type interaction and thereby
   the chemical potential is shifted by an amount $G_2 \rho$, 
  $ \mu = \mu_r + G_2 \rho $ \cite{rf:AsaYaz,rf:KitKoiKunNem}.
This graph Fig.2 has the peculiarity; the line of the graph is lowered in the range
   $ 370 \; {\rm MeV}  \stackrel{<}{\sim}  \mu  \stackrel{<}{\sim}  470  \; {\rm MeV} $.
This peculiarity will be discussed later.
 In Fig.3, the nontrivial solution of $U_A$ is figured, and one finds that the quark spin
   polarization occurs in the chemical potential range 
    $ 370  \; {\rm MeV}  \stackrel{<}{\sim}  \mu  \stackrel{<}{\sim}  470  \; {\rm MeV} $.
It is notable that the spin polarization disappears when the chemical potential becomes too large,
    $ \mu  \stackrel{>}{\sim}  470  \; {\rm MeV} $,
   and this point will be also discussed in the later subsection.

Next, when $ m=5 \, {\rm MeV} $, we find by numerical calculations that
    it is necessary to satisfy
   $ G_2/G_1  \stackrel{>}{\sim} 6.3 $  in order to occur the spin polarization $U_A >0$.
One can see that the necessary value $ G_2/G_1  \stackrel{>}{\sim} 6.3 $ in the case of
    $ m=5 \, {\rm MeV} $ is smaller than the value $ G_2/G_1  \stackrel{>}{\sim} 7.37 $
   in the chiral limit case $m=0$.
Although our NJL model of one flavor at finite density differs from the NJL model of
   three flavor in the vacuum state, the above value $ G_2/G_1  \stackrel{>}{\sim} 6.3 $
   is fairly larger than the value $ G_2/G_1 \sim 0.7 $ which value is expected in the NJL model
   of three flavor in the vacuum state \cite{rf:KliLutVogWei}.
We then take another large value of $m$ and find the value $ G_2/G_1 $ necessary
   for obtaining spin polarization $U_A >0$.

Finally, when the current mass is taken to be $ m=130 \, {\rm MeV} $, we find that
    it is necessary to satisfy
   $ G_2/G_1  \stackrel{>}{\sim} 1.89 $  so as to occur the spin polarization $U_A >0$.
%
%
%
%
\begin{figure}
  \begin{center}
     \includegraphics[height=7cm]{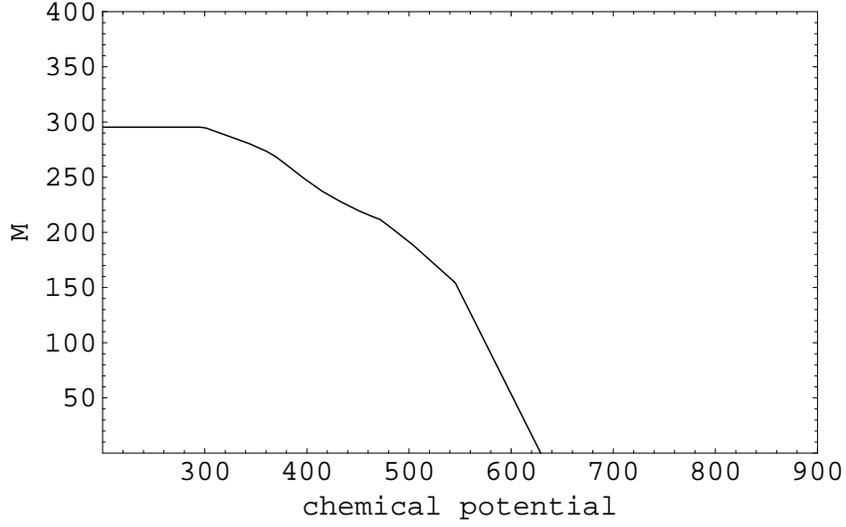}
  \end{center}
   \caption{Dynamical quark mass $M({\rm MeV})$ as a function of
                  the chemical potential $\mu\,({\rm MeV})$
                  when the ratio is $G_2/G_1 = 7.37 $
                  and the current quark mass is $m=0$.
                  The $M$ in Fig.2 and the nontrivial solution $U_A$ in Fig.3 are the solution of the
                  simultaneous equations
                   Eq.(\ref{ci}), Eq.(\ref{ck}), and Eq.(\ref{co}).}
\label{fig:2}
\end{figure}
%
%
%
%
%
%
\begin{figure}
    \begin{center}
       \includegraphics[height=7cm]{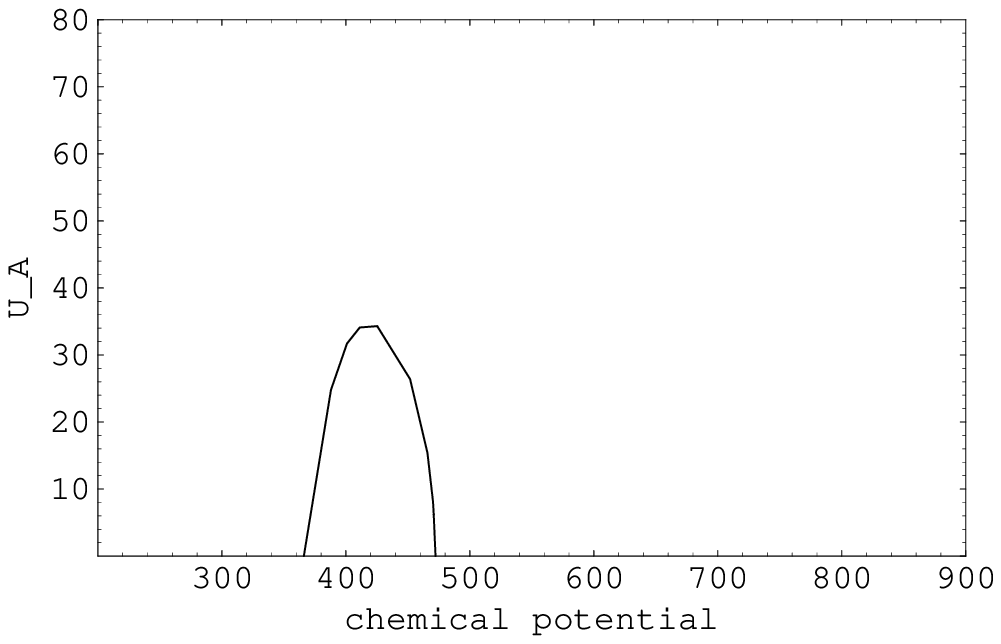}
    \end{center}
    \caption{ Spin polarization $U_A({\rm MeV})$ as a function of
                  the chemical potential $\mu\,({\rm MeV})$
                  when the ratio is $G_2/G_1 = 7.37 $
                  and the current quark mass is $m=0$.
                  The $M$ in Fig.2 and the nontrivial solution $U_A$ in Fig.3 are the solution of the
                  simultaneous equations
                    Eq.(\ref{ci}), Eq.(\ref{ck}), and Eq.(\ref{co}). }
\label{fig:3}
\end{figure} 
%
%
%
%
%
\begin{figure}
  \begin{center}
     \includegraphics[height=7cm]{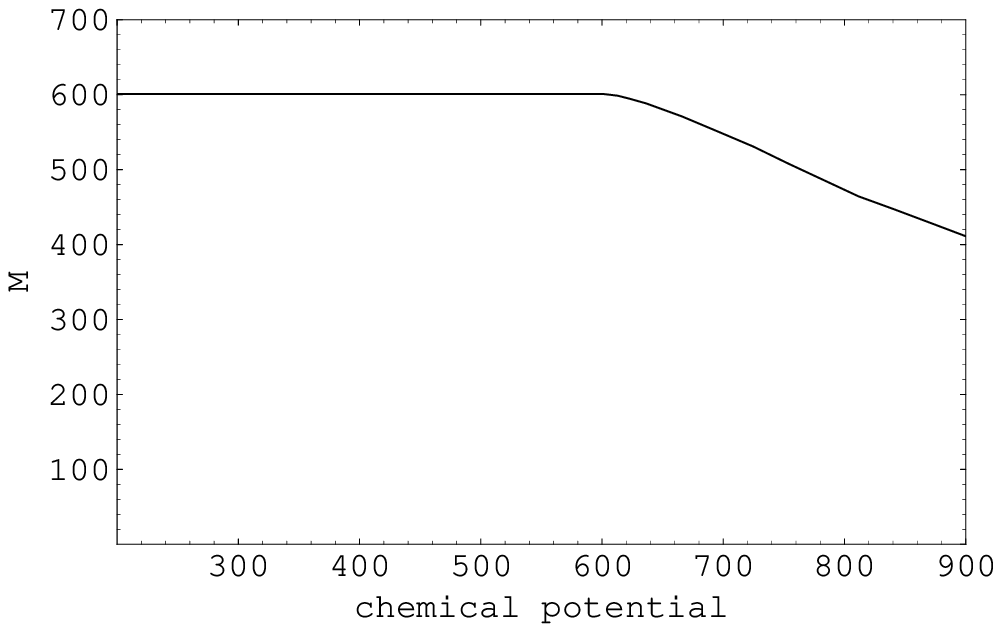}
  \end{center}
   \caption{Dynamical quark mass $M({\rm MeV})$ as a function of
                  the chemical potential $\mu\,({\rm MeV})$
                  when the ratio is $G_2/G_1 = 1.89 $
                  and the current quark mass is $m= 130 \, {\rm MeV} $.
                  The $M$ in Fig.4 and the nontrivial solution $U_A$ in Fig.5 are the solution of the
                  simultaneous equations
                   Eq.(\ref{ci}), Eq.(\ref{ck}), and Eq.(\ref{co}).}
\label{fig:4}
\end{figure}
%
%
%
%
%
%
\begin{figure}
    \begin{center}
       \includegraphics[height=7cm]{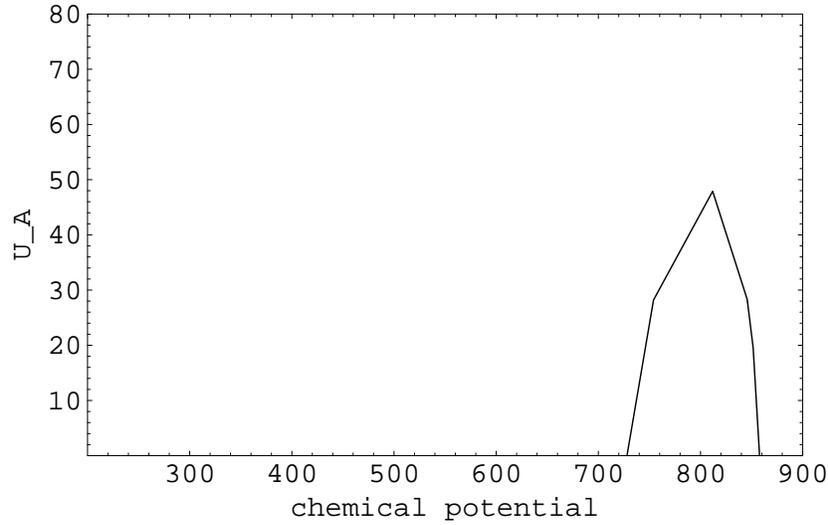}
    \end{center}
    \caption{ Spin polarization $U_A({\rm MeV})$ as a function of
                  the chemical potential $\mu\,({\rm MeV})$
                  when the ratio is $G_2/G_1 = 1.89 $
                  and the current quark mass is  $ m=130 \, {\rm MeV} $.
                  The $M$ in Fig.4 and the nontrivial solution $U_A$ in Fig.5 are the solution of the
                  simultaneous equations
                    Eq.(\ref{ci}), Eq.(\ref{ck}), and Eq.(\ref{co}). }
\label{fig:5}
\end{figure} 
While $ G_2/G_1  \stackrel{>}{\sim} 1.89 $ is still larger than the value $0.7$,
   we understand that the necessary
   value $ G_2/G_1 $ for obtaining spin polarization becomes smaller
   if the current quark mass $m$ becomes larger.
The numerical results for $ m=130 \, {\rm MeV} $ and $ G_2/G_1 = 1.89 $ are represented
   in Fig.4 and Fig.5.
The value of $\mu$ at which the dynamical mass $M$ begins to decrease in Fig.4 is larger
   than that value of $\mu$ in Fig.2, because the current mass $m$ has larger value 
   $ 130 \, {\rm MeV} $ in Fig.4 than in Fig.2.
When the current mass $m$ is $ 130 \, {\rm MeV} $ 
   (constituent mass $M_0 =601\, {\rm MeV} $), the spin polarization occurs in the
   chemical potential range 
   $ 730 \, {\rm MeV} \,  \stackrel{<}{\sim}  \mu  \stackrel{<}{\sim}  860 \, {\rm MeV} $
   as seen from Fig.5.
When, on the other hand, $m$ is $0$ (constituent mass $M_0 =295\, {\rm MeV} $), 
   the spin polarization occurs in the
   range 
   $ 370 \, {\rm MeV} \,  \stackrel{<}{\sim}  \mu  \stackrel{<}{\sim}  470 \, {\rm MeV} $
   as seen from Fig.3.
The difference between these two ranges of $\mu$ in Fig.5 and Fig.3 comes from
   the fact that the chemical potential $\mu$ must exceed the value of the constituent mass
   $M_0$ so that the quark number density $\rho$ has a positive value.
%
%
%
\subsection{Effect of the spin polarization $U_A$ on the dynamical quark mass $M$,
         and vice versa. }
We have solved numerically the simultaneous self consistent equations for $M$, $ \mu_r$,
   and $U_A$
   when the current mass  $m= 0, \, 5  \, {\rm MeV}, \, 130 \, {\rm MeV}$, respectively.
Since these equations are the simultaneous ones, $U_A$ has an effect on $M$,
   and vice versa.
In this subsection, we study the effect of $U_A$ on $M$ and the effect of $M$
   on $U_A$ through the case of $m=0$.
The behavior of the dynamical mass $M$ in Fig.2 is influenced by the spin polarization $U_A$,
   and the behavior of $U_A$ in Fig.3 is influenced by the dynamical mass $M$.

First, let us consider the effect of the spin polarization $U_A$ on the dynamical mass $M$.
As mentioned before, the behavior of $M$ in Fig.2 has the peculiarity that the line of the 
   graph is lowered in the range
    $ 370  \; {\rm MeV} \stackrel{<}{\sim}  \mu  \stackrel{<}{\sim}  470  \; {\rm MeV} $.
This peculiarity will be caused by the influence of the spin polarization $U_A$ being
   non-zero in that range of $\mu$ as seen in Fig.3.
In order to confirm this, we solve numerically the self consistent equation for $M$ and that
   for $\mu_r$ with setting $U_A =0$ in the whole range of $\mu$,
   $ 0 \le \mu \le \Lambda$.
The numerical result is shown with the dotted line in Fig.6, and the graph of Fig.2 is also
   shown in Fig.6 with the solid line for comparison.
When the solid line is compared with the dotted line in Fig.6, we find that if the quark spin
   polarization occurs $U_A>0$, it operates to lower the dynamical quark mass $M$
   which is generated by spontaneous symmetry breakdown of chiral symmetry.

Next, let us consider the effect of the dynamical mass $M$ on the spin polarization $U_A$.
As seen in Fig.2, $M$ is a monotone decreasing function of $\mu$, which passes through
   a point $ ( \mu, M )=( 452 \; {\rm MeV} , 219 \; {\rm MeV} )$.
As mentioned before, the behavior of $U_A$ in Fig.3 has the peculiarity that the spin
   polarization $U_A$ disappears when the chemical potential becomes too large,
   $ \mu  \stackrel{>}{\sim} 470 \; {\rm MeV} $.
This peculiarity may be caused by the fact that the dynamical mass $M$ is the monotone
   decreasing function of $\mu$.
%
%
%
%
%
\begin{figure}
    \begin{center}
      \includegraphics[height=7cm]{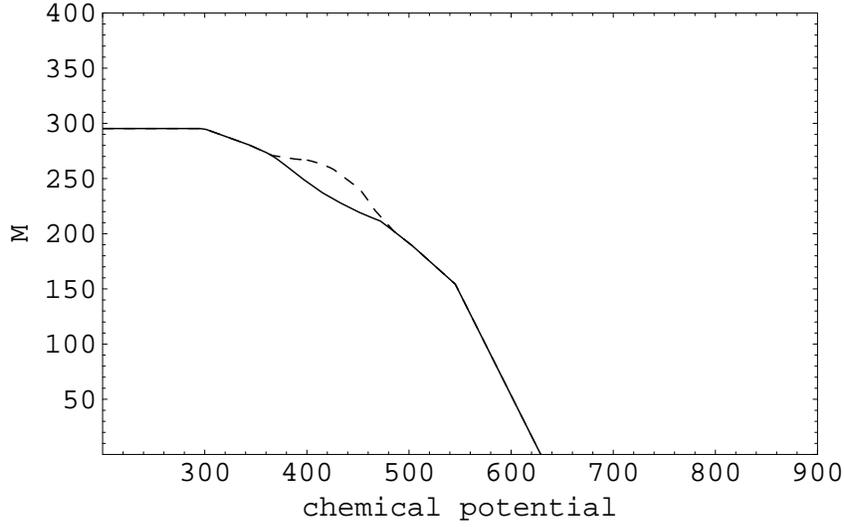}
  \end{center}
   \caption{ Dynamical quark mass $M({\rm MeV})$ as a function of
                  the chemical potential $\mu\,({\rm MeV})$
                  when the ratio is $G_2/G_1 = 7.37 $
                  and the current quark mass is  $ m=0 $.
                  The dotted line represents the dynamical mass when there is no spin polarization
                  ( $U_A=0$ ).
                  The solid line  represents the dynamical mass with the nontrivial solution $U_A$
                  (the same result shown in Fig.2). }
\label{fig:6}
\end{figure}
%
%
%
%
%
%
\begin{figure}
   \begin{center}
      \includegraphics[height=7cm]{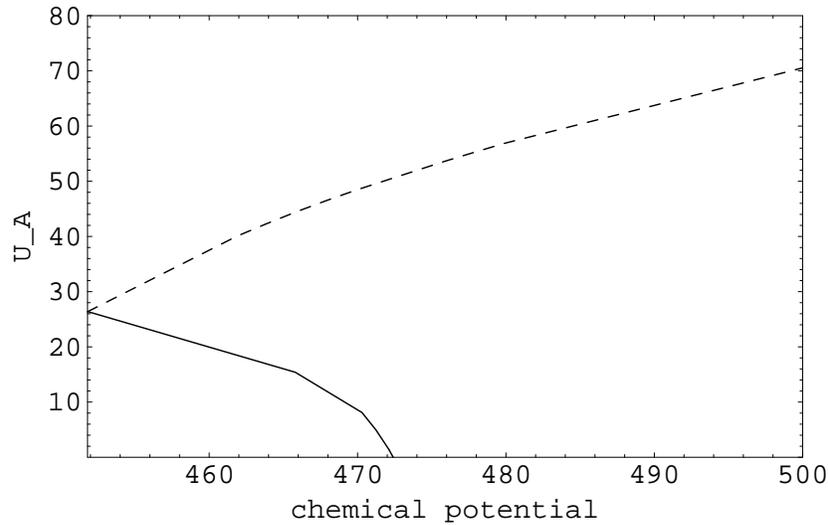}
   \end{center}
   \caption{ Spin polarization $U_A({\rm MeV})$ as a function of
                   the chemical potential $\mu\,({\rm MeV})$
                  when the ratio is $G_2/G_1 = 7.37 $.
                  The solid line is a part ($ \, \mu \ge  452 \; {\rm MeV}  \, $)
                   of the nontrivial solution shown in Fig.3.
                  The dotted line is obtained when we set the dynamical mass
                   $M= {\rm constant} =219 \; {\rm MeV} $.  }
\label{fig:7}
\end{figure}
In order to confirm this, we examine how the $U_A$ behaves when the $M$ is taken to be 
   a constant.
Specifically speaking, we solve numerically the self consistent equation for $U_A$ and
   that for $\mu_r$ with setting $M= {\rm constant} =219 \; {\rm MeV} $
   \footnote{ The $M= {\rm constant} =219 \; {\rm MeV} $ is not a solution of
           the self consistent equation for $M$. }
   in the range of $ \mu \ge 452 \; {\rm MeV} $.
The numerical result is shown with the dotted line in Fig.7, which shows that the spin polarization
   $U_A$ becomes larger as $\mu$ increases even in the range of large 
   $ \mu  \stackrel{>}{\sim} 470 \; {\rm MeV} $ if the $M$ is constant
   ( $ 219 \; {\rm MeV} $).
In Fig.7, a part of the graph of Fig.3 ( $ \mu \ge 452 \; {\rm MeV} $ )
   is also shown with the solid line.
By comparing this solid line with the dotted line in Fig.7, we find that, when the chiral
   symmetry is gradually restored as the chemical potential $\mu$ increases
     ( $ \mu \ge 452 \; {\rm MeV} $ ), the
   dynamical quark mass $M$ reduces
   ($M \le 219 \; {\rm MeV} $, see Fig.2)
   and this reduction of the dynamical quark mass $M$ operates
   to suppress the spin polarization $U_A$
    (see the solid line in Fig.7).

So far we have argued the special case of $m=0$, yet the qualitative properties of the
   effect of $U_A$ on $M$ and the effect of $M$ on $U_A$ are the same with 
   the cases of $m= 5  \, {\rm MeV} $ and $ 130  \, {\rm MeV} $.
%
%
%
%
%
%
\section{Conclusion}
We studied the relation between the quark spin polarization and spontaneous chiral symmetry
   breaking at finite density by use of the NJL type model with the number of colors
   $ N_c =3 $ and  the number of flavors  $ N_f =1 $, which includes interactions in vector
   and axial-vector channel.
The self consistent equations for chiral condensate $ \langle \bar \psi \psi \rangle $,
   spin polarization $ \langle  \psi^{\dag} \Sigma_z \psi \rangle $, and quark number
   density $ \langle \bar \psi  \gamma^0 \psi \rangle $ are derived respectively in the 
   framework of the Hartree approximation (mean field approximation).
These simultaneous equations are solved numerically with
   regarding the quark current mass $m$ and the coupling constant ratio $G_2/G_1$
   as free parameters. 
Consequently, we find that in the flavor $N_f =1$ model the spin polarization is possible
   at finite density if the ratio $G_2/G_1$ is larger than $O(1)$.
When the current quark mass $m$ becomes heavier, the necessary value $G_2/G_1$ for
   obtaining the spin polarization becomes smaller.
For example, if $ m= 130  \, {\rm MeV} $, the necessary value is 
   $ G_2/G_1  \stackrel{>}{\sim} 1.89 $.

The spin polarization occurs in the following way.
The spin polarization begins to occur at a certain value of the chemical potential, at which
   the dynamical quark mass is still a few hundred  $ \, {\rm MeV}  $.
The reason why the spin polarization occurs are that the quark number density is finite
   and that the dynamical quark mass remains relatively large value about a few
   hundred  $ \, {\rm MeV}  $.
If the effective mass $M$ of the quark is taken to be
   a small value such as
  $ 5  \; {\rm MeV} $, the spin polarization
   does not take place.
When one increases the chemical potential furthermore, the spin polarization becomes weaker
   and finally disappears, where the chiral symmetry is gradually restored and the dynamical
   mass $M$ is reduced from its value in the vacuum state.
The reason why the spin polarization disappears at the large chemical potential is that the 
   dynamical quark mass is reduced by restoring gradually the chiral symmetry.
The effects of the dynamical mass $M$ on the spin polarization $U_A$ are discussed above;
   conversely, what is the effect of the spin polarization on the dynamical mass?
The numerical calculations show that, if the spin polarization occurs $U_A>0$, it operates to
   lower the dynamical quark mass $M$ which is generated by spontaneous chiral
   symmetry breaking.

In studying the quark spin polarization with the moderate density at which chiral symmetry
   is partially broken spontaneously, it is necessary to consider the relation between the
   spin polarization and spontaneous chiral symmetry breaking \cite{rf:NakMarTat};
   the present paper 
   attempts to investigate such issue.
However, there remains several problems not discussed here, and we show below;
\begin{itemize}
  \item It is necessary to make sure that the nontrivial solution $U_A$ of the self consistent
            equations gives the minimum value of the thermodynamical potential.
  \item The effect of quark confinement is not considered, since the NJL model we use is
            an effective theory that does not confine quarks.
            We therefore treat the system at finite density described by the NJL model
            as quark matter in the present paper.
            Although the lattice QCD simulation
             demonstrates
            quark confinement, it is difficult to apply it to the system with finite density. The 
            development of this field is anticipated.
  \item We need to study the case of $SU(2)$ or $SU(3)$ flavors in which the condition of
            charge neutrality is imposed.
            According to the NJL model analysis in the vacuum state, the ratio $G_2/G_1$
             is taken to be $G_2/G_1\sim 0.7$ when the number of flavor is three.
            In our one flavor model at finite density, it is necessary to satisfy 
             $ G_2/G_1  \stackrel{>}{\sim} 1.89 $ in order to occur spin polarization
            if the current quark mass is $ m= 130  \, {\rm MeV} $.
             We therefore need to study the spin polarization around 
             $G_2/G_1 \sim O(1)$ in the case of three flavors.
  \item If the color superconductivity is considered, one needs to investigate the relationship
             among the spin polarization, spontaneous chiral symmetry breaking, and the
             the color superconductivity.
  \item We have done the numerical calculations with $T=0$. It would be interesting to study 
             numerically the finite temperature case $T>0$.
\end{itemize}
%
%
%
%
%
%
%
\section*{Acknowledgments}

The author thanks the Yukawa Institute for Theoretical Physics at Kyoto University.
Discussions during the YITP workshop YITP-W-06-07
   on "Thermal Quantum Field Theories and Their Applications" were useful to 
   complete this work.
%
%
%
%
\newpage 
\noindent{\Large\bf Appendix}
\appendix 
%
%
\section{The calculation of the propagator $G_A$.}
\renewcommand{\theequation}{A.\arabic{equation}}
\setcounter{equation}{0}
The $ G_A $, Eq.(\ref{bh}), is calculated as follows.
Following the notation used in the textbook of Bjorken and Drell \cite{rf:BjoDre},
   we write
\begin{equation}
    \left[ \, p \!\!\!/ - M \,  + U_{ A} \, \gamma_5 \gamma^3  \right] 
   =   \left( \begin{array}{cc}
             p_0-U_A \sigma_3-M    &   -  \mbox{\boldmath $ p $} 
                                                                           \cdot   \mbox{\boldmath $ \sigma $}    \cr 
             \mbox{\boldmath $ p $}  \cdot   \mbox{\boldmath $ \sigma $}
                                                                                            &   -p_0+U_A \sigma_3-M    
                       \end{array}   \right).
  \label{pa}
\end{equation}
By the way, the inverse of the matrix
\begin{equation}
  X =  \left( \begin{array}{cc}
               A  &  B  \cr
               C  &  D   
                       \end{array}   \right),
  \label{pb}
\end{equation}
can be obtained by\cite{rf:WesBag}
\begin{equation}
  X^{-1} =  \left( \begin{array}{cc}
               -C^{-1} D U   &   Y  \cr
               U                   &   -B^{-1} A Y
                       \end{array}   \right),
  \label{pc}
\end{equation}
where 
\begin{eqnarray}
    U &\equiv& ( B-A C^{-1} D )^{-1},    \nonumber  \\
   Y &\equiv& (C-D B^{-1} A )^{-1},   
  \label{pd}
\end{eqnarray}
provided $B^{-1}$ and $C^{-1}$ exist.
Using this formula, we get Eq.(\ref{bh}).
%
%
%
%
%
%
\section{Summation over the Matsubara frequency}
\renewcommand{\theequation}{B.\arabic{equation}}
\setcounter{equation}{0}
Let us use the standard technique \cite{rf:Kap,rf:LeB} to calculate the sum 
   $\sum_{j=-\infty}^{\infty} $ over the Matsubara frequency
   $\omega_j = ( 2 j +1 ) \pi T $.
Let $g(p^0)$ be a function of $p^0$ which decreases faster than $1/p^0$
   for  $ \vert p_0 \vert \rightarrow \infty $.
Putting $p_0 = i \omega_j +\mu_r$, we obtain the following equation by the contour
   integration \cite{rf:LeB},
\begin{equation}
  {1 \over \beta} \sum_{j=-\infty}^{\infty} g ( p_0 = i \omega_j +\mu_r )
  = \sum \frac{ {\rm Res} \, g( p_0 ) }{e^{\beta ( p_0 -\mu_r ) } +1 }.
  \label{qa}
\end{equation}
Now, in Eq.(\ref{cg}), we take as follows
\begin{eqnarray}
  g(p_0 ) &=&  \frac{ 4 \, M  ( p^2-M^2+U_A^2 \, ) }
                           { ( p^2-M^2-U_A^2 \, )^2 - 4 \, U_A^2 \, ( M^2 + p_z^2 ) }    \nonumber \\
    &=&  \frac{ 4 \, M  ( p^2-M^2+U_A^2 \, ) }{ 2 \, ( \epsilon_2^2-\epsilon_1^2 ) } \sum_{n=1}^{4}
           \frac{ (-1 )^n }{ \epsilon_n } \frac{1}{ ( p_0 - \epsilon_n ) },
  \label{qb}
\end{eqnarray}
then Eq.(\ref{ch}) is derived by the formula Eq.(\ref{qa}).
%
%
%
%
%
\section{Explicit expressions for the self consistent equations}
\renewcommand{\theequation}{C.\arabic{equation}}
\setcounter{equation}{0}
The self consistent equations Eq.(\ref{ci}), Eq.(\ref{ck}), and Eq.(\ref{co}) with $T=0$
   can be calculated analytically when $ 0 \le U_A<M$.
The results are
\begin{eqnarray}
  M &=&  m                            \nonumber  \\
  & & \hskip-0.6cm   - G_1 \frac{N_c}{ (2 \pi )^2 }  M  \biggl[ 
        \biggl\{ ( \mu_{\rm r} - 3 U_A ) \sqrt{ ( \mu_{\rm r} - U_A )^2 -M^2 }          \nonumber  \\
   & & \hskip-0.3cm      + ( 2 \, U_A (\mu_{\rm r} - U_A )-M^2 ) \log  
        \frac{  ( \mu_{\rm r} - U_A )+ \sqrt{ ( \mu_{\rm r} - U_A )^2 -M^2 } }{M} \biggr\}
        \theta (  \mu_{\rm r} - ( M+U_A) )        \nonumber  \\
  & &  \hskip-0.3cm      +\biggl\{ ( \mu_{\rm r} + 3 U_A ) \sqrt{ ( \mu_{\rm r} + U_A )^2 -M^2 }        
\nonumber  \\
   & & \hskip-0.3cm      + ( - 2 \, U_A (\mu_{\rm r} + U_A )-M^2 ) \log  
        \frac{  ( \mu_{\rm r} + U_A )+ \sqrt{ ( \mu_{\rm r} + U_A )^2 -M^2 } }{M} \biggr\}
        \theta (  \mu_{\rm r} - ( M - U_A) )       \biggr]       \nonumber  \\
   & & \hskip-0.6cm     + G_1  \frac{N_c}{ (2 \pi )^2 }  M  \biggl[ ~
         \mu_{\rm r} \rightarrow \sqrt{ \Lambda^2 + M^2 } ~ \biggr],
  \label{ra}
\end{eqnarray}
\vskip0.5cm
\begin{eqnarray}
   \hskip-0.5cm \mu_{\rm r} &=&  \mu                              \nonumber  \\
  & &  \hskip-0.6cm     -G_2  \frac{N_c}{ (2 \pi )^2 } \biggl[ ~
        \biggl\{  ~ \frac{1}{3} \,  \left( - 2 M^2 + (  \mu_{\rm r} - U_A ) 
        ( 2 \, \mu_{\rm r} + U_A ) \right)  \sqrt{ ( \mu_{\rm r} - U_A )^2 -M^2 }          \nonumber  \\
   & & \hskip1cm      - U_A  M^2  \log  
        \frac{  ( \mu_{\rm r} - U_A )+ \sqrt{ ( \mu_{\rm r} - U_A )^2 -M^2 } }{M} \biggr\}
        \theta (  \mu_{\rm r} - ( M+U_A) )   \biggr]      \nonumber  \\
  & &  \hskip-0.6cm     - G_2  \frac{N_c}{ (2 \pi )^2 } \biggl[ ~ U_A \rightarrow -U_A ~ \biggr],
  \label{rb}
\end{eqnarray}
\vskip0.5cm
\begin{eqnarray}
 U_A   &=&  G_2 \frac{N_c}{ (2 \pi )^2 } \biggl[ ~
        \biggl\{  ~ \frac{1}{3} \,  \left( 4 M^2 - (  \mu_{\rm r} - U_A ) 
        ( \mu_{\rm r} + 2 \,U_A ) \right)  \sqrt{ ( \mu_{\rm r} - U_A )^2 -M^2 }          \nonumber  \\
   & & \hskip1cm      - ( \mu_{\rm r} - 2 \, U_A ) M^2  \log  
        \frac{  ( \mu_{\rm r} - U_A )+ \sqrt{ ( \mu_{\rm r} - U_A )^2 -M^2 } }{M} \biggr\}
        \theta (  \mu_{\rm r} - ( M+U_A) )   \biggr]      \nonumber  \\
  & &  \hskip-0.1cm      -  G_2  \frac{N_c}{ (2 \pi )^2 } \biggl[ ~ U_A \rightarrow -U_A ~ \biggr].
  \label{rc}
\end{eqnarray}
\vskip 2cm
%
%
%
%
%
%

%
%
%
\end{document}